\documentclass{PoS}

\pdfoutput=1

\usepackage{amsmath,amssymb}

\newcommand{\Tr}{\operatorname{Tr}}

\title{Lattice Landau gauge via Stereographic
  Projection\footnotetext{Supported by the Australian Research Council.}}

\ShortTitle{Lattice Landau gauge via Stereographic Projection}

\author{\speaker{Lorenz von Smekal}\!,
  \  Alexander Jorkowski, Dhagash Mehta and Andr\'e Sternbeck\\
 Centre for the Subatomic Structure of Matter, School of
  Chemistry and Physics,\\
 The University of Adelaide, SA 5005, Australia\\
        E-mail: \parbox[t]{11cm}{\email{lorenz.smekal@adelaide.edu.au},\\
          \email{alexander.jorkowski@student.adelaide.edu.au},
          \email{dhagash.mehta@adelaide.edu.au}, 
          \email{andre.sternbeck@adelaide.edu.au}}   
}

\abstract{The complete cancellation of Gribov copies and the Neuberger
  $0/0$ problem of lattice BRST can be avoided in modified lattice Landau
gauge. In compact $U(1)$, where the problem is a lattice artifact,
there remain to be Gribov copies but their number is exponentially
reduced. Moreover, there is no cancellation of copies there as the
sign of the Faddeev-Popov determinant is positive. Applied to the
maximal Abelian subgroup this avoids the perfect cancellation amongst
the remaining Gribov copies for $SU(N)$ also. In addition, based on a
definition of gauge fields on the lattice as
stereographically-projected link variables, it provides a framework
for gauge fixed Monte-Carlo simulations. This will include all Gribov
copies in the spirit of BRST. Their average is not zero, as
demonstrated explicitly in simple models. This might resolve present
discrepancies between gauge-fixed lattice and continuum studies of QCD
Green's functions.} 

\FullConference{8th Conference Quark Confinement and the Hadron Spectrum \\
		 September 1-6, 2008\\
		 Mainz, Germany}

\begin{document}


\vspace*{-1.4cm}

\section{Introduction} 
\vspace{-.1cm}

The Green's functions of QCD provide a basis for hadron phenomenology
\cite{Alkofer:2000wg}.  Their infrared behaviour 
is also known to contain information about the realisation
of confinement in Landau gauge QCD.
Dyson-Schwinger equation (DSE) studies
\cite{vonSmekal:1997is} have established  
 that the gluon propagator alone does not provide
long-range interactions of a strength sufficient to confine quarks, and
that the infrared dominant correlations are instead mediated by the
Faddeev-Popov ghosts of this formulation, whose propagator was found
to be infrared enhanced. 
This can be understood in terms of confinement in QCD
\cite{Alkofer:2000wg,Alkofer:2001iw,Lerche:2002ep}, as 
a consequence of the Kugo-Ojima criterion which is based on the
realisation of the unfixed global gauge symmetries of the covariant
continuum formulation. In order to distinguish confinement
from Coulomb and Higgs phases it requires: (a) The massless single particle
singularity in the transverse gluon correlations of perturbation
theory must be screened non-perturbatively to avoid long-range fields
and charged superselection sectors as in QED. (b) The global gauge
charges must remain well-defined and unbroken to avoid the Higgs
mechanism. In Landau gauge, in which the (Euclidean) 
gluon and ghost propagators 
are parametrised by the two invariant functions $Z$ and $G$,
respectively, this entails that

\vspace{-.3cm}
\begin{equation}
\mbox{(a):} \;\; \lim_{p^2\to 0} Z(p^2)/p^2 \, < \,\infty \; ; \qquad
\mbox{(b):} \;\; \lim_{p^2\to 0} G^{-1}(p^2) \, =  \, 0 \; . 
\label{KO-IR}
\end{equation}
The translation of (b) into the infrared enhancement of the ghost
propagator thereby rests on the ghost/anti-ghost symmetry of the
Landau gauge or the symmetric Curci-Ferrari gauges. 
It represents an additional boundary condition
on DSE solutions which then lead to the prediction of a conformal
infrared behaviour for the gluonic correlations in Landau gauge QCD
\cite{Lerche:2002ep}. 
In fact, this behaviour is directly tied to the validity and
applicability of the framework of local quantum field theory for
non-Abelian gauge theories beyond perturbation theory. The subsequent
verification of this infrared behaviour with a variety of different
functional methods in the continuum meant a remarkable success. 
These methods which all lead to the same prediction 
include studies of their Dyson-Schwinger Equations (DSEs)
\cite{Lerche:2002ep}, Stochastic Quantisation \cite{Zwanziger:2001kw},
and of the Functional Renormalisation Group Equations
(FRGEs)~\cite{Pawlowski:2003hq}.  
This prediction amounts to infrared asymptotic forms

\vspace{-.8cm}
\begin{equation}
   \label{infrared-gh_gl}
  Z(p^2) \, \sim\, (p^2/\Lambda^2_{\mbox{\tiny QCD}})^{2\kappa_Z} \; , \;\; 
  \; \mbox{and} \;\;  G(p^2) \, \sim \, 
  (p^2/\Lambda_{\mbox{\tiny QCD}}^2)^{-\kappa_G} \; ,
\end{equation}
for $p^2 \to 0$, which are both determined by a unique critical
infrared exponent $ \kappa_Z = \kappa_G \equiv \kappa $,
with $ 0.5 < \kappa < 1$. Under a mild regularity assumption
on the ghost-gluon vertex \cite{Lerche:2002ep}, the value of this
exponent is furthermore obtained as $\kappa \, = \, (93 -
\sqrt{1201})/98 \, \approx \, 0.595 $ 
\cite{Lerche:2002ep,Zwanziger:2001kw}. 

The conformal nature of this infrared behaviour in the pure Yang-Mills
sector of Landau gauge QCD is evident in the generalisation to
arbitrary gluonic correlations \cite{Alkofer:2004it} which has 
furthermore been shown to represent a unique {\em scaling solution}
 \cite{Fischer:2006vf}. In particular,
the ghost-gluon vertex is then infrared finite and the
non-perturbative running coupling of
\cite{vonSmekal:1997is} approaches an infrared
fixed-point,

\vspace{-.25cm}
\begin{equation} 
 \alpha_S(p^2) \, = \, \frac{g^2}{4\pi} Z(p^2) G^2(p^2)
\, \to \, \alpha_c \quad \mbox{for} \;\; p^2 \to 0 \; . 
\label{alpha_minimom}
\end{equation}

\vspace{-.15cm}

\noindent
If the ghost-gluon vertex is regular at $p^2 =0$, its value is
maximised and given by $\alpha_c \, \approx
  \, {8.9}/{N_c}\, $ \cite{Lerche:2002ep}. 
However, the uniqueness of infrared scaling does not rule out
solutions with infrared finite gluon propagator and  a  
ghost propagator with a free massless-particle
singularity, {\it i.e.},
$Z(p^2) \, \sim \, p^2/M^2\, $, and $G(p^2) \, \sim \,
\mathrm{const.} $,
for $p^2 \to 0$. This solution corresponds to $\kappa_Z = 1/2 $ and $
\kappa_G = 0$. It does not satisfy the scaling relation $\kappa_Z
=\kappa_G $ in   (\ref{infrared-gh_gl}) because transverse gluons decouple
for momenta $p^2 \ll M^2 $, and it is therefore called the {\em
  decoupling solution} \cite{Fischer:2008uz}.  
The interpretation of 
(\ref{alpha_minimom}) as a running coupling does not make sense in the
infrared in this case, in which there is no infrared fixed-point and
no conformal infrared behaviour.   


Because of the inevitable finite-volume effects, 
early lattice studies of the gluon and ghost propagators 
could have been consistent with both, the scaling solution or the
decoupling solution. Recently, the finite-volume effects have been
analysed carefully in the Dyson-Schwinger equations to demonstrate how
the scaling solution is approached in the infinite volume limit there
\cite{Fischer:2007pf}. This is clearly not what is being observed,
however, in more recent $SU(2)$ lattice data on impressively large lattices
\cite{Sternbeck:2007ug,Cucchieri:2007md}.
Present lattice data is fully consistent with the decoupling solution
which poses the question whether we are perhaps comparing apples with
oranges when comparing the minimal lattice Landau gauge
correlations with those of local quantum field theory in the infrared? 

The latter is based on a cohomology construction of a
physical Hilbert space over the indefinite metric spaces of covariant
gauge theory from the representations of the Becchi-Rouet-Stora-Tyutin
(BRST) symmetry. But do we have a non-perturbative definition of a
BRST charge in presence of Gribov copies?  In the most direct translation
of BRST symmetry on the lattice, there is a perfect cancellation among
these gauge copies which gives rise to the famous Neuberger $0/0$
problem. It asserts that the expectation value of any gauge invariant
(and thus physical) observable in a lattice BRST formulation will
always be of the indefinite form $0/0$ \cite{Neuberger1987} 
and therefore prevented such formulations for more than 20 years now. 
In present lattice implementations of the Landau gauge this problem is
avoided because the numerical procedures are based on minimisations of
a gauge-fixing potential w.r.t.~gauge transformations. To find
absolute minima is not feasible on large lattices as this is a
non-polynomially hard computational problem. One therefore settles for
local minima which in one way or another, depending on the algorithm,
samples gauge copies of the first Gribov region among which there is
no cancellation. For the same reason, however, this is not a BRST
formulation. The emergence of the decoupling solution can thus not be
used to dismiss the Kugo-Ojima criterion of covariant gauge theory in
the continuum.

\vspace{-3mm}

\section{Lattice BRST and the Neuberger $0/0$ Problem}

\vspace{-1.5mm}

In principle, a BRST symmetry could be implemented on the lattice by
inserting the partition function of a topological model with
BRST exact action into the gauge invariant lattice measure. Because of
its topological nature, this gauge-fixing partition function
$Z_{\mbox{\tiny GF}}$ will be independent of gauge orbit and gauge
parameter. The problem is that in the standard formulation this
partition function calculates the Euler characteristic $\chi $ 
of the lattice gauge group which vanishes \cite{Schaden1998},

\vspace{-.4cm}
\begin{equation}
Z_{\mbox{\tiny GF}} = \chi(SU(N)^{\#{\rm sites}}) =
\chi(SU(N))^{\#{\rm sites}} = 0^{\# {\rm sites}}\; . 
\label{EulerC}
\end{equation}

\vspace{-.15cm}

\noindent
Neuberger's $0/0$ problem of lattice BRST arises because we have then
inserted zero instead of unity (according to the Faddeev-Popov
prescription) into the measure of lattice gauge theory.
On a finite lattice, such a topological model is equivalent to a
problem of supersymmetric quantum mechanics with Witten index 
$ Z_{\mbox{\tiny GF}}$, except that for gauge-fixing we need a 
model with non-vanishing Witten index to avoid the Neuberger $0/0$
problem. 
Then however, just as the supersymmetry of the corresponding quantum
mechanical model, such a lattice BRST cannot break.

In Landau gauge the Neuberger zero,
$Z_{\mbox{\tiny GF}} = 0$, arises from the perfect cancellation of
Gribov copies via the Poincar\'e-Hopf theorem. The gauge-fixing potential
for a generic link configuration thereby plays the role of a Morse
potential for gauge transformations and the Gribov copies are its
critical points (the global gauge transformations need to remain
unfixed so that there are strictly speaking only $(\# $sites$-1)$
factors of $\chi(SU(N)) = 0$ in (\ref{EulerC})). The Morse inequalities
then immediately imply that there are at least $2^{(N-1)(\# {\rm
    sites}-1)}$ such copies in $SU(N)$ on the lattice, or $2^{\# {\rm
    sites}-1}$ in compact $U(1)$, and equally many with either sign
of the
Faddeev-Popov determinant. 

The topological origin of the zero originally observed by Neuberger in a
certain parameter limit due to uncompensated Grassmann ghost
integrations in standard Faddeev-Popov theory \cite{Neuberger1987} 
becomes particularly
evident in the ghost/anti-ghost symmetric Curci-Ferrari gauge with its
quartic ghost self-interactions.  In this gauge the same
parameter limit leads to computing the zero in (\ref{EulerC}) from a
product of independent Gauss-Bonnet integral expressions,
for each site of the lattice \cite{Sme08}, corresponding to the Gauss-Bonnet
limit of the equivalent supersymmetric quantum mechanics model in
which only constant paths contribute \cite{Birmingham1991}.
The indeterminate form of physical observables
as a consequence of (\ref{EulerC}) can be regulated by a
Curci-Ferrari mass term. While such a mass $m$ decontracts the double
BRST/anti-BRST algebra, which is known to result in a loss of unitarity,
observables can then be meaningfully defined in the limit  
$m \to 0$ via l'Hospital's rule \cite{Sme08}.

\vspace{-3mm}

\section{Lattice Landau Gauge from Stereographic Projection}

\vspace{-1mm}

The 0/0 problem due to the vanishing Euler characteristic of $SU(N)$  
is avoided when fixing the gauge only up to the maximal Abelian
subgroup $U(1)^{N-1}$ because the Euler characteristic of the coset
manifold is non-zero. The corresponding lattice BRST has been
explicitly constructed for $SU(2)$ \cite{Schaden1998}, where the coset
manifold is the 2-sphere and $\chi(SU(2)/U(1)) = \chi(S^2) = 2$. 
This indicates that the Neuberger problem might be solved when
that of compact $U(1)$ is, where the same cancellation of lattice Gribov
copies arises because $\chi(S^1) = 0$. A surprisingly simple
solution to this problem is possible, however, by stereographically 
projecting the circle $S^1 \to \mathbb{R}$ which can be achieved by a
modification of the minimising potential
\cite{vonSmekal:2007ns}.  The resulting potential is convex to the
above and leads to a positive definite Faddeev-Popov operator for
compact $U(1)$ where there is thus no cancellation of Gribov
copies, but $Z_{\mbox{\tiny GF}}^{U(1)} = N_{\mbox{\tiny GC}}$, for
$N_{\mbox{\tiny GC}}$ Gribov copies, which follows from a simple example
of a Nicolai map \cite{Birmingham1991}.  
As compared to the standard lattice Landau gauge the number of copies 
is furthermore exponentially reduced. This is easily
verified explicitly in low dimensional models.
While $N_{\mbox{\tiny GC}}$ grows exponentially with the number of sites in
the standard case as expected, the stereographically projected version
has only $N_{\mbox{\tiny GC}} = N_x $ copies on a periodic chain of length
$N_x$,   and $\ln N_{\mbox{\tiny GC}} \sim N_t \ln N_x $ on a 
$2D$ lattice of size $N_t \times  N_x$ in Coulomb gauge, for example, 
and in both cases
their number is verified to be independent of the gauge orbit. 

Applying the same techniques to the maximal Abelian subgroup
$U(1)^{N-1}$, the generalisation to $SU(N) $ lattice gauge theories is
possible when the odd-dimensional spheres $S^{2n+1} $, $n=1,\dots
N\!-\!1$, of its parameter space are stereographically projected to
$\mathbb{R} \times \mathbb{R}P(2n)$. In absence of the cancellation of
the lattice artifact Gribov copies along the $U(1)$ circles, the
remaining cancel\-lations between copies of either sign in $SU(N)$,
which will persist in the continuum limit, are then necessarily
incomplete, however, because $\chi(\mathbb{R}P(2n)) = 1$. 
For $SU(2)$ this program is straightforward. Starting from a modified
 gauge-fixing potential \cite{vonSmekal:2007ns} one defines
 stereographically-projected gauge fields on 
%
the lattice (see \cite{Sternbeck:2008wg}), 

\vspace{-.4cm}
\[
{A}_{x\mu} = \frac{1}{2ia}\left(\widetilde{U}_{x\mu} -
    \widetilde{U}^{\dagger}_{x\mu}\right)
  \; , \;\; \text{where}\quad
  \widetilde{U}_{x\mu} 
     \equiv \frac{2 U_{x\mu}}{1+ \frac{1}{2} \Tr U_{x\mu}} \; , 
  \label{gauge-fields}
\]
such that the gauge-fixing condition is given by the usual lattice
divergence but in terms of these stereographically-projected gauge
fields. A particular advantage of the non-compact gauge fields is
that they allow to resolve the gauge condition of the 
stereographically-projected lattice Landau gauge by Hodge
decomposition. This provides a framework for gauge-fixed Monte-Carlo
simulations which is currently being developed for $SU(2)$ in 2 dimensions
\cite{Felix}. In the low-dimensional models mentioned above it can
furthermore be verified explicitly that the corresponding 
gauge-fixing partition function is indeed given by 
$ Z_{\mbox{\tiny GF}}^{SU(2)} 
   \, = \,  Z_{\mbox{\tiny GF}}^{U(1)}  \, \not= \, 0   \, $, 
as expected from $\chi(\mathbb{R}P(2)) = 1$. 


\vspace{-1mm}

\section{Conclusions and Outlook}

\vspace{-1mm}

Comparisons of the infrared behaviour of QCD Green's functions as
obtained from lattice Landau gauge implementations based on
minimisations of a gauge-fixing potential and from continuum studies
based on BRST symmetry have to be taken with a grain of salt.  
Evidence of the asymptotic conformal behaviour predicted by the latter
is seen in the strong coupling limit of lattice Landau gauge where
such a behaviour can be observed at large lattice momenta $a^2p^2\gg
1$ \cite{Sternbeck:2008wg}. 
Observed deviations from scaling at small momenta in the
strong-coupling limit 
are not finite-volume effects,
but discretisation dependent and hint at a breakdown of BRST symmetry
arguments beyond perturbation theory in this
approach. Non-perturbative lattice BRST has been plagued by the
Neuberger $0/0$ problem, but its improved topological understanding provides
ways to overcome this problem. The most promising one at this point
rests on stereographic projection to define gauge fields on the
lattice together with a modified lattice Landau gauge. This new
definition has the appealing feature that it will allow
gauge-fixed Monte-Carlo simulations in close analogy to the continuum
BRST methods which it will thereby elevate to a non-perturbative
level.

\vspace{-2mm}



\begin{thebibliography}{99}\itemsep -1mm

\bibitem{Alkofer:2000wg}
  R.~Alkofer and L.~von Smekal,
  Phys.\ Rept.\  {\bf 353} (2001) 281.

\bibitem{vonSmekal:1997is}
  L.~von Smekal, A.~Hauck and R.~Alkofer,
  Phys.\ Rev.\ Lett.\  {\bf 79} (1997) 3591;
  Annals Phys.\  {\bf 267} (1998) 1.

\bibitem{Alkofer:2001iw}
  R.~Alkofer, L.~von Smekal and P.~Watson,
  arXiv:hep-ph/0105142.


\bibitem{Lerche:2002ep}
  Ch.~Lerche and L.~von Smekal,
  Phys.\ Rev.\  D {\bf 65} (2002) 125006.

\bibitem{Zwanziger:2001kw}
  D.~Zwanziger,
  Phys.\ Rev.\  D {\bf 65} (2002) 094039.

\bibitem{Pawlowski:2003hq}
  J.~M.~Pawlowski, D.~F.~Litim, S.~Nedelko and L.~von Smekal,
  Phys.\ Rev.\ Lett.\  {\bf 93} (2004) 152002.

\bibitem{Alkofer:2004it}
  R.~Alkofer, C.~S.~Fischer and F.~J.~Llanes-Estrada,
  Phys.\ Lett.\  B {\bf 611} (2005) 279.


\bibitem{Fischer:2006vf}
  C.~S.~Fischer and J.~M.~Pawlowski,
  Phys.\ Rev.\  D {\bf 75} (2007) 025012.

\bibitem{Fischer:2008uz}
  C.~S.~Fischer, A.~Maas and J.~M.~Pawlowski,
  arXiv:0810.1987 [hep-ph].


\bibitem{Fischer:2007pf}
  C.~S.~Fischer, A.~Maas, J.~M.~Pawlowski and L.~von Smekal,
  Annals Phys.\  {\bf 322} (2007) 2916.

\bibitem{Sternbeck:2007ug}
  A.~Sternbeck, L.~von Smekal, D.~B.~Leinweber and A.~G.~Williams,
  PoS {\bf LAT2007} (2007) 340.

\bibitem{Cucchieri:2007md}
  A.~Cucchieri and T.~Mendes,
  PoS {\bf LAT2007} (2007) 297.

\bibitem{Neuberger1987}
  H.~Neuberger, Phys.\ Lett.\ B {\bf 175} (1986) 69; {\it ibid.}
  {\bf 183} (1987) 337.

\bibitem{Schaden1998}
  M.~Schaden, {Phys. Rev.} \textbf{D59} (1998) 014508.

\bibitem{Sme08}
  L.~von Smekal, M.~Ghiotti and A.~G.~Williams,
  Phys.\ Rev.\  D {\bf 78} (2008) 085016.

\bibitem{Birmingham1991}
  D.~Birmingham, M.~Blau, M.~Rakowski, G.~Thompson, Phys.\ Rept.\
  {\bf 209} (1991) 129.

\bibitem{vonSmekal:2007ns}
  L.~von Smekal, D.~Mehta, A.~Sternbeck, A.~G.~Williams,
  PoS {\bf LAT2007} (2007) 382.

\bibitem{Sternbeck:2008wg}
  A.~Sternbeck and L.~von Smekal, PoS {\bf LATTICE2008} (2008) 267;
  and arXiv:0811.4300 [hep-lat];\\[-3pt] see also
  A.~Sternbeck's talk at this conference. 

\bibitem{Felix} A.~Jorkowski, Honours thesis, University of Adelaide,
  November 2008;\\[-3pt] 
   A.~Jorkowski, A.~Sternbeck
  and L.~von Smekal, in  preparation. 
\end{thebibliography}
\end{document}